\documentclass[twocolumn,showpacs,preprintnumbers,amsmath,amssymb]{revtex4}
\usepackage{graphicx}% Include figure files
\usepackage{dcolumn}% Align table columns on decimal point
\usepackage{bm}% bold math
\usepackage{epstopdf}
\usepackage{color}
\def\ga{\,\,\raise0.14em\hbox{$>$}\kern-0.76em\lower0.28em\hbox{$\sim$}\,\,}
\def\la{\,\,\raise0.14em\hbox{$<$}\kern-0.76em\lower0.28em\hbox{$\sim$}\,\,}

\begin{document}

\preprint{PRC}

\title{Systematical studies of the E1 photon strength functions combining  Skyrme-HFB+QRPA model and experimental giant dipole resonance properties}

\author{Y. Xu$^{1}$, S. Goriely$^{2}$, E. Khan$^3$}

\affiliation{$^{1}$Extreme Light Infrastructure - Nuclear Physics (ELI-NP), Horia Hulubei National Institute for R\&D in Physics and Nuclear Engineering (IFIN-HH), 077125 Buchurest-Magurele, Romania}
\affiliation{$^{2}$Institut d'Astronomie et d'Astrophysique, Universite Libre de Bruxelles, CP-226, 1050 Brussels, Belgium}
\affiliation{$^3$ Institut de Physique Nucl\'eaire, Universit\'e Paris-Sud, IN2P3-CNRS,
Universit\'e Paris-Saclay, F-91406 Orsay Cedex, France}
\date{\today}

\begin{abstract}
Valuable theoretical predictions of nuclear dipole excitations in the whole nuclear chart are of great interest for
different applications, including in particular nuclear astrophysics. We present here the systematic study of the electric dipole (E1) photon strength functions combining the microscopic Hartree-Fock-Bogoliubov plus Quasiparticle Random Phase Approximation (HFB+QRPA) model and the parametrizations constrained by the available experimental giant dipole resonance (GDR) data. For about 10000 nuclei with 8 $\leq$ Z $\leq$ 124 lying between the proton and the neutron drip-lines on nuclear chart, the particle-hole ($ph$) strength distributions are computed using the HFB+QRPA model under the assumption of spherical symmetry and making use of the BSk27 Skyrme effective interaction derived from the most accurate HFB mass model (HFB-27) so far achieved. Large-scale calculations of the BSk27+QRPA E1 photon strength functions are performed in the framework of a specific folding procedure describing the damping of nuclear collective motion empirically. In particular, three phenomenological improvements are considered in this folding procedure. First, two interference factors are introduced and adjusted to reproduce at best the available experimental GDR data. Second, an empirical expression accounting for the deformation effect is applied to describe the peak splitting of the strength function. Third, the width of the strength function is corrected by a temperature-dependent term, which effectively increases the de-excitation photon strength function at low-energy. The theoretical E1 photon strength functions as well as the extracted GDR peaks and widths are comprehensively compared with available experimental data. A relatively good agreement with data indicates the reliability of the present calculations. Eventually, the astrophysical rates of (n,$\gamma$) reactions for all the 10000 nuclei with 8 $\leq$ Z $\leq$ 124 lying between the proton and the neutron drip-lines are estimated using the present E1 photon strength functions. The resulting reaction rates are compared with the previous BSk7+QRPA results as well as the Gogny-HFB+QRPA predictions based on the D1M interaction.
\end{abstract}

\pacs{24.30.Cz, 21.60.Jz, 25.60.Tv, 26.30.-k}

%\keywords{GDR, HFB, QRPA, E1-strength, Neutron Capture, Nucleosynthesis}%Use showkeys class option if keyword
                              %display desired

\maketitle

\section{Introduction}
\label{sect_intro}

About half of the nuclei with $A > 60$ observed in nature are formed by the rapid neutron-capture process (r-process) occurring in explosive stellar events \cite{gognyref1}. The r-process is believed to take place in environments characterized by high neutron density ($N_{n} > 10^{20} {\rm cm}^{-3}$), so that successive neutron captures proceed into neutron-rich regions well off the $\beta$-stability valley forming exotic nuclei that cannot be produced and therefore studied in the laboratory. When the temperature or the neutron density required for the r-process are low enough to break the equilibrium of (n,$\gamma$)-($\gamma$,n), the distribution of the r-process abundance depends directly on the neutron capture rates of the so-produced exotic neutron-rich nuclei \cite{gognyref2}. The neutron capture rates are commonly evaluated within the framework of the statistical model of Hauser-Feshbach, although the direct capture contribution may play an important role for very exotic neutron-rich nuclei \cite{gognyref3}. The Hauser-Feshbach model makes the fundamental assumption that the capture process takes place with the intermediary formation of a compound nucleus in thermodynamic equilibrium. In this approach, the Maxwellian-averaged neutron capture rate at temperatures of relevance in r-process environments strongly depends on the electromagnetic interaction, {\it i.e.} on the photon de-excitation probability. Therefore, a reliable extrapolation of the photon strength functions out towards the neutron-drip line is required for a proper description of the r-process and prediction of the resulting abundance distribution.

In massive star, both during the hydrostatic and explosive burning phases, as well as in type-Ia supernovae, high temperatures of a few billion degrees can also be at the origin of heavy-element nucleosynthesis by the so-called p-process \cite{Arnould03}. The p-process is responsible for the galactic production of the 35 neutron-deficient stable nuclei and mainly involves photoreactions on pre-existing nuclei synthesized by the neutron-capture processes. It is driven by ($\gamma$,n), ($\gamma$,p)  and ($\gamma$,$\alpha$) reactions as well as their reverse reactions mainly on neutron-deficient nuclei. Here also,  rates are determined within the Hauser-Feshbach model where the photon strength function plays a fundamental role for estimating both the photoabsorption reaction rates and their reverse radiative capture rates  \cite{Arnould03}.

Large-scale calculations of dipole photon strength function have traditionally been performed on the basis of the phenomenological Lorentzian model \cite{Goriely19,RIPL3}. However, this approach suffers from shortcomings of various sorts. First, it is unable to predict the enhancement of the dipole photon strength at energies around and below the neutron separation energy demonstrated by various experiments \cite{Goriely19}. Second, even if a Lorentzian-like function provides a suitable representation of the dipole strength for stable nuclei, the location of its maximum and its width remain to be predicted from some systematics or underlying model for each nucleus. The phenomenological Lorentzian model consequently lacks reliability when dealing with exotic nuclei that are particularly important to the astrophysical applications. On the other hand, the reliability of the E1 strength predictions can be greatly improved by the use of microscopic (or semi-microscopic) models. Provided satisfactory reproduction of available experimental data, the more microscopic the underlying theory, the greater the confidence in the extrapolations out towards the experimentally unreachable regions. However,  microscopic approaches are rarely used for large-scale predictions of the E1 strength function, mainly because the fine tuning required to reproduce accurately a large experimental data set is very delicate and time-consuming. The prominent exceptions are represented by Refs. \cite{NPA2002,BSk7,Daoutidis12,Gogny2016} where the E1 photon strength functions for the whole nuclear chart were derived within the mean-field plus quasiparticle random-phase approximation (QRPA) approach.

In the present paper, a new effort is made to estimate systematically the E1 strength function on the basis of the non-relativistic Hartree-Fock-Bogoliubov plus QRPA (HFB+QRPA) approach, in a way similar to our previous study \cite{BSk7}. However, since its publication, significant improvements have been achieved, essentially in the determination of Skyrme interactions.
In particular, a new effective standard Skyrme interaction, labeled as BSk27, was derived together with the HFB-27 nuclear mass model, leading to the most accurate mass model  ever achieved within the framework of the nuclear energy density functional theory \cite{hfb27}. This effective interaction has not been tested yet on QRPA estimates of giant resonances. Furthermore, compared to the previous HFB+QRPA study based on the BSk7 Skyrme interaction \cite{BSk7}, a new fitting procedure aiming at reproducing experimental photoabsorption data has been implemented to constrain the parameters entering in the QRPA calculations and the corrections needed beyond the one particle - one hole (1p-1h) QRPA. In the framework of HFB+QRPA approach with the recent BSk27 Skyrme interaction, the present study aims to predict a complete set of E1 photon strength functions within the whole nuclear chart, simultaneously reproducing the available experimental data to date. It is expected that these treatments can significantly improve the predictions of E1 strength and consequently of the neutron capture rates. It also allows us to test the systematic uncertainties affecting the prediction of the photon strength function for exotic neutron-rich nuclei, in particular by comparing it to the recently determined D1M+QRPA predictions based on the axially deformed Gogny-HFB plus QRPA calculations with the D1M Gogny interaction \cite{Gogny2016,Goriely19}. The derivation of the M1 photon strength function within the same framework is postponed to a future study.

The paper is organized as follows. In Sec.~II, the QRPA formalism is sketched and the folding prescription (the damping method for collective motions) to derive the continuous E1 photon strength function from the QRPA strength distributions is described. The large-scale calculations of the E1 photon strength functions are performed in Sec.~III, taking into account: (1) the determination of the parameters in the folding prescription based on experimental constraints, (2) the impact of nuclear deformation, and (3) the nuclear temperature correction. In Sec.~IV, the neutron capture reaction rates for astrophysical applications are correspondingly computed based on the present E1 photon strength functions predicted by the HFB plus QRPA model with the BSk27 Skyrme force. A summary is given in Sec.~V.

\section{Theory}
\label{HFBQRPA}

\subsection{HFB plus QRPA model}

The HFB plus QRPA method allows to investigate, in a self-consistent way, the nuclear structure properties of the ground state as well as collective excitations for the nuclei ranging from the valley of stability to the drip-line. The QRPA considers nuclear excitation as a collective superposition of two quasiparticle (qp) states built on top of the HFB ground state and this collective aspect of the excitation makes the QRPA an accurate tool to study the E1 photon strength function in both closed and open shell nuclei \cite{PRingbook}.

The ground state properties used here are derived from the HFB-27 mass model \cite{hfb27} obtained within the HFB framework. HFB-27 is the most accurate mass model we ever achieved within the framework of the nuclear energy density functional theory. It is characterized by a model root mean square deviation $\sigma_{mod}$ = 0.503 MeV with respect to all the 2408 available mass data \cite{Wang17} for nuclei with neutron and proton numbers larger than 8. The  Skyrme effective interaction, labeled as BSk27, corresponds to the conventional form of a 10-parameter Skyrme force with a 4-parameters $\delta$-function pairing force treated in the Bogoliubov framework.  In addition, as determined by realistic calculations and by experiments, the underlying Skyrme functional yields a realistic description of infinite homogeneous nuclear matter properties, specifically including the incompressibility coefficient, the pressure in charge-symmetric nuclear matter, the neutron-proton effective mass splitting, the stability against spin and spin-isospin fluctuations, as well as the neutron-matter equation of state. Such properties play a key role in the description of giant resonances \cite{Harakeh01}. All details can be found in Refs.~\cite{hfb27,hfb1,hfb2,hfb3,hfb4,hfb5,Chamel15}.

The QRPA calculation of the E1 photon strength function is performed on top of the HFB ground state. The derivation of the QRPA response is detailed in Ref.~\cite{BSk7} using the Green's function formalism. The QRPA response is obtained from the time-dependent HFB equations \cite{PRingbook}
\begin{eqnarray}
i\hbar\frac{\partial\mathcal{R}}{\partial t} = [\mathcal{H}(t)+\mathcal{F}(t),\mathcal{R}(t)],
\label{e1}
\end{eqnarray}
in which $\mathcal{R}$ is the time-dependent generalized density, $\mathcal{H}$ is the HFB Hamiltonian, and $\mathcal{F}$ is the external oscillating field. In the small amplitude limit the time-dependent HFB equations become
\begin{eqnarray}
\hbar \omega\mathcal{R}' = [\mathcal{H}',\mathcal{R}^{0}] + [\mathcal{H}^{0},\mathcal{R}'] + [F,\mathcal{R}^{0}]
\label{e2}
\end{eqnarray}
where $'$ stands for the perturbed quantity.
The variation of the generalized density $\mathcal{R}'$ is expressed as a column vector estimated from $\bm{\rho'}$, the transpose of $(\rho'~\kappa'~\bar{\kappa}')$.
%Only the change of the particle-hole (ph) density $\rho'$ is needed to know in the variance with the RPA, while the variation of the three basic quantities ($\rho'$, $\kappa'$ and $\bar{\kappa}'$ in the column vector $\mathcal{R}'$) should be calculated in QRPA.
In the three dimensional space, the first dimension represents the $ph$ subspace, the second the particle-particle ($pp$) one, and the third the hole-hole ($hh$) one.
%%Therefore, the response matrix has 3$\times$3=9 coupled elements in QRPA, but only one in RPA.

The variation of the HFB Hamiltonian can be expressed in terms of the second derivatives of the HFB energy functional $\mathcal{E}$[$\rho$, $\kappa$, $\bar{\kappa}$] with respect to the densities
\begin{eqnarray}
\mathbf{H}' = \mathbf{V}\bm{\rho}',
\label{e3}
\end{eqnarray}
where the residual interaction matrix $\mathbf{V}$ is written as
\begin{eqnarray}
\mathbf{V}^{ab}(\mathbf{r}\sigma,\mathbf{r}'\sigma')=\frac{\partial^{2}\mathcal{E}}{\partial\bm{\rho}_{b}(\mathbf{r}'\sigma')\partial\bm{\rho}_{\bar{a}}(\mathbf{r}\sigma)}
\label{e4}
\end{eqnarray}
with $a,b=1,2,3$. Here, the notation $\bar{a}$ means that whenever $a$ is 2 or 3, then $\bar{a}$ is 3 or 2.

The relevant quantity is the QRPA Green's function $\mathbf{G}$, which relates the perturbing external field to the density change by
\begin{eqnarray}
\bm{\rho}' = \mathbf{G}\mathbf{F}.
\label{e5}
\end{eqnarray}
Replacing Eqs.~(\ref{e3}), (\ref{e4}) and (\ref{e5}) in Eq.~(\ref{e2}), yields the Bethe-Salpeter equation
\begin{eqnarray}
\mathbf{G} = (1-\mathbf{G}_{0}\mathbf{V})^{-1}\mathbf{G}_{0}=\mathbf{G}_{0}+\mathbf{G}_{0}\mathbf{V}\mathbf{G}
\label{e6}
\end{eqnarray}
corresponding to a set of 3 $\times$ 3 = 9 coupled equations. The unperturbed Green's function $\mathbf{G}_{0}$ is defined by
\begin{eqnarray}
\mathbf{G}_{0}^{ab}(\mathbf{r}\sigma,\mathbf{r}'\sigma';\omega) &=& \sum_{ij}\ \Big[ \frac{\mathcal{U}_{ij}^{a1}(\mathbf{r}\sigma)\bar{\mathcal{U}}_{ij}^{*b1}(\mathbf{r}'\sigma')}{\hbar\omega-(E_{i}+E_{j})+i\eta} \nonumber\\
&&-\frac{\mathcal{U}_{ij}^{a2}(\mathbf{r}\sigma)\bar{\mathcal{U}}_{ij}^{*b2}(\mathbf{r}'\sigma')}{\hbar\omega+(E_{i}+E_{j})+i\eta}\Big],
\label{e7}
\end{eqnarray}
where $E_{i}$ and $E_{j}$ are the energies of single qp state and ${\mathcal{U}}_{ij}$ are the 3 $\times$ 2 matrices with the elements calculated from the HFB wave functions $U$ and $V$ \cite{qrpaodd,BSk7}.

The HFB wave functions $U$ and $V$ are obtained by solving the HFB equations in the HF basis with $\Psi_{i}^{HF}=\sum_{k}D_{ik}\chi_{k}$ expanded on an oscillator basis, which reads
\begin{eqnarray}
U_{l}(\mathbf{r})=\sum_{i,E_{i}>0}U_{li}\Psi_{i}^{HF}(\mathbf{r})=\sum_{k}[\sum_{i,E_{i}>0}U_{li}D_{ik}]\chi_{k}(\mathbf{r})
\label{e8}
\end{eqnarray}
and
\begin{eqnarray}
V_{l}(\mathbf{r})=\sum_{i,E_{i}>0}V_{li}\Psi_{i}^{HF}(\mathbf{r})=\sum_{k}[\sum_{i,E_{i}>0}V_{li}D_{ik}]\chi_{k}(\mathbf{r}).
\label{e9}
\end{eqnarray}
In Eqs.~(\ref{e8}) and (\ref{e9}), $U_{li}$ and $V_{li}$ are the coefficients of the Bogoliubov transformation. The prescription for the calculations of the HFB wave functions can be found in Ref. \cite{hfb1}.

For the transitions from the ground state to excited states within the same nucleus, only the ($ph$,$ph$) component of $\mathbf{G}$ plays a role, and the strength is thus given by
\begin{eqnarray}
S(\omega)=-\frac{1}{\pi}Im\int F^{11*}(\mathbf{r})\mathbf{G}^{11}(\mathbf{r},\mathbf{r}';\omega)F^{11}(\mathbf{r}')d\mathbf{r}d\mathbf{r}'.
\label{e10}
\end{eqnarray}
Following our previous work \cite{BSk7}, we restrict ourselves to perform the present calculations of the QRPA strength distribution S($\omega$) by assuming the spherical symmetry. The residual interaction corresponding to the velocity-dependent terms of the Skyrme force used for the HFB calculation is approximated in the ($ph$,$ph$) subspace by its Landau-Migdal limit \cite{residual}. On the other hand, the ($pp$, $pp$) part of the residual interaction is self-consistently derived from the pairing force \cite{hfb1}.

Based on the ground state properties derived by BSk27 interaction and the above-described QRPA formalism, the BSk27 $ph$ QRPA strength distributions S($\omega$) have been calculated for about 10000 nuclei with $8 \leq Z \leq 124$. The computations are performed up to a maximum transition energy $\omega_{max}$ = 30 MeV with a step of 0.1 MeV. All the qp states up to an energy cutoff of 60 MeV are included. The spurious center-of-mass state should come out at zero energy in a fully consistent calculation. Due to the adoption of the interaction in Landau-Migdal form, the consistency between mean field and residual qp interaction is broken and the isoscalar $J^{\pi}$ = $1^{-}$ spurious state becomes imaginary. Therefore, to remedy this defect, the residual interaction is renormalized by a factor $\alpha$ on the nuclei of interest. The typical value of $\alpha$ in the [0.85,1] range drives the spurious state to zero energy. To calculate the photon strength function of odd-$A$ and odd-odd nuclei, the procedure developed in Ref.~\cite{hfb1,qrpaodd} is adopted.
%This procedure was shown to give rather satisfactory results \cite{Gogny2016}.

\begin{figure}
\hskip -0.6cm
\includegraphics[width=9cm]{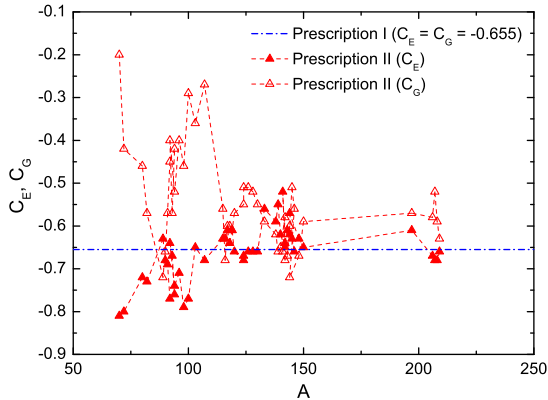}
\vskip -0.3cm
\caption{\label{fig1-cecg} Interference factors $C_{E}$ and $C_{G}$ for each of the 48 spherical nuclei obtained from prescription I (constant value $C_{E}$ = $C_{G}$ = -0.655 illustrated by dash-dot line) and prescription II (different values of $C_{E}$ and $C_{G}$ shown by triangles connected with dash lines) as a function of the atomic mass.}
\end{figure}

\begin{figure*}
\hskip -0.6cm
\includegraphics[width=17cm]{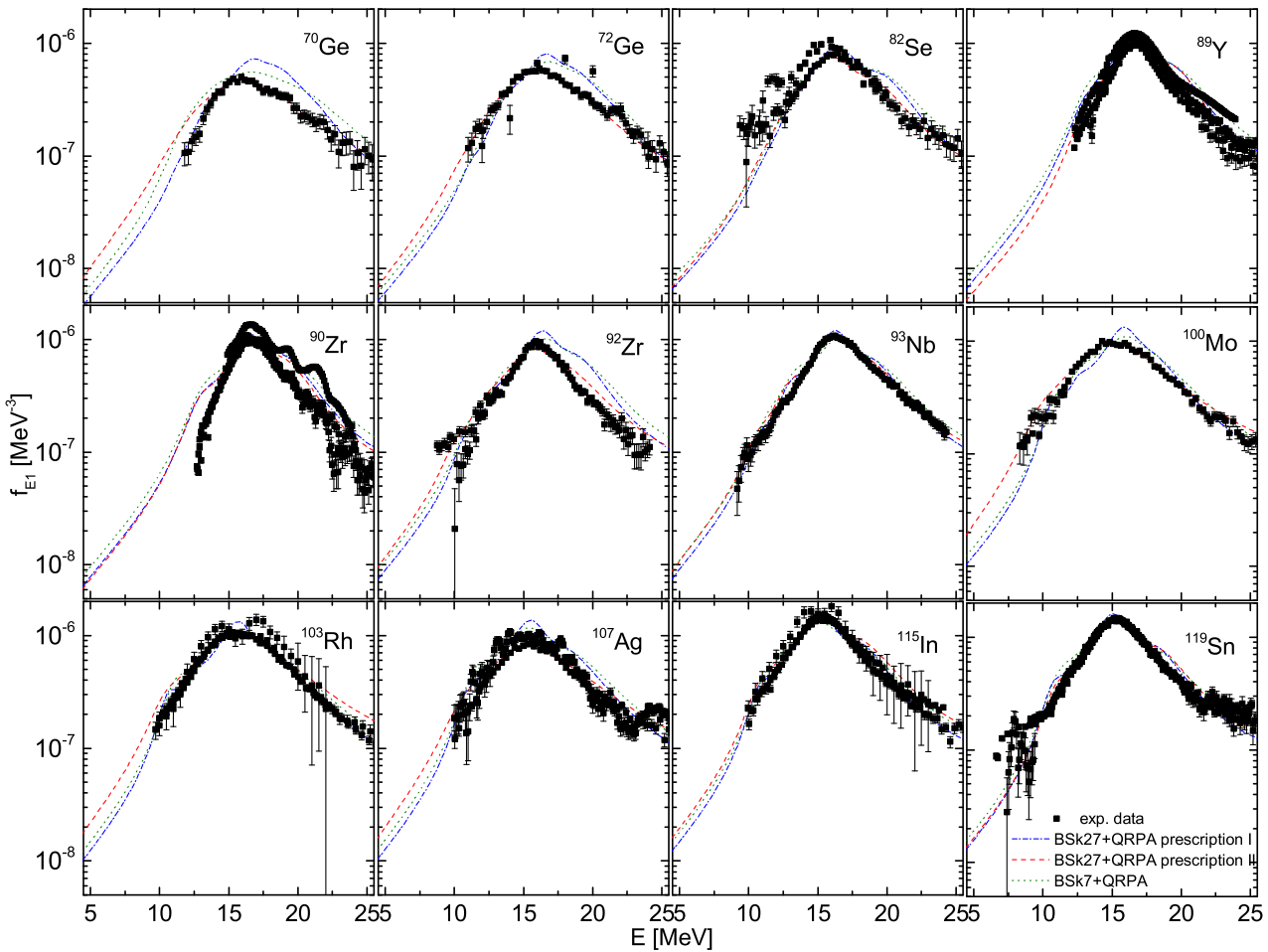}
\vskip -0.3cm
\caption{\label{fig2-abs} Comparisons between the photon strength function extracted from experimental photoabsorption cross sections \cite{Goriely19} and the 2 sets of BSk27+QRPA photon strength functions calculated with the interference factors $C_{E}$ and $C_{G}$ of prescription I (orange dash-dot-dot line) and prescription II (red dashed line) for 12 representative spherical nuclei. The previous BSk7+QRPA predictions (green dotted line) are also shown for  comparison.}
\end{figure*}

\begin{figure*}
\hskip -0.6cm
\includegraphics[width=17cm]{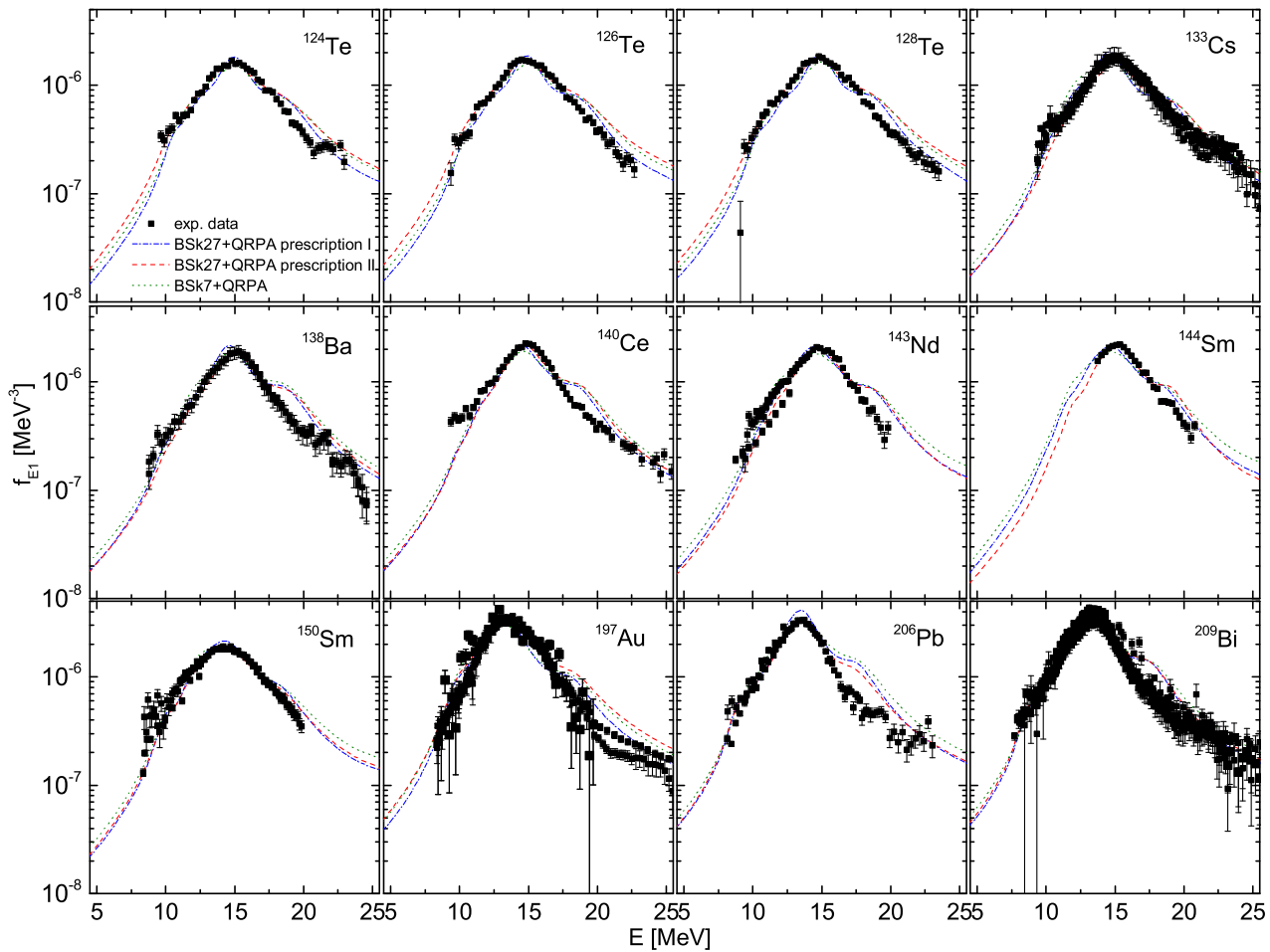}
\vskip -0.3cm
\caption{\label{fig3-abs} Same as Fig.~\ref{fig2-abs} for another 12 representative spherical nuclei.}
\end{figure*}

\begin{figure*}
\hskip -0.6cm
\includegraphics[width=17cm]{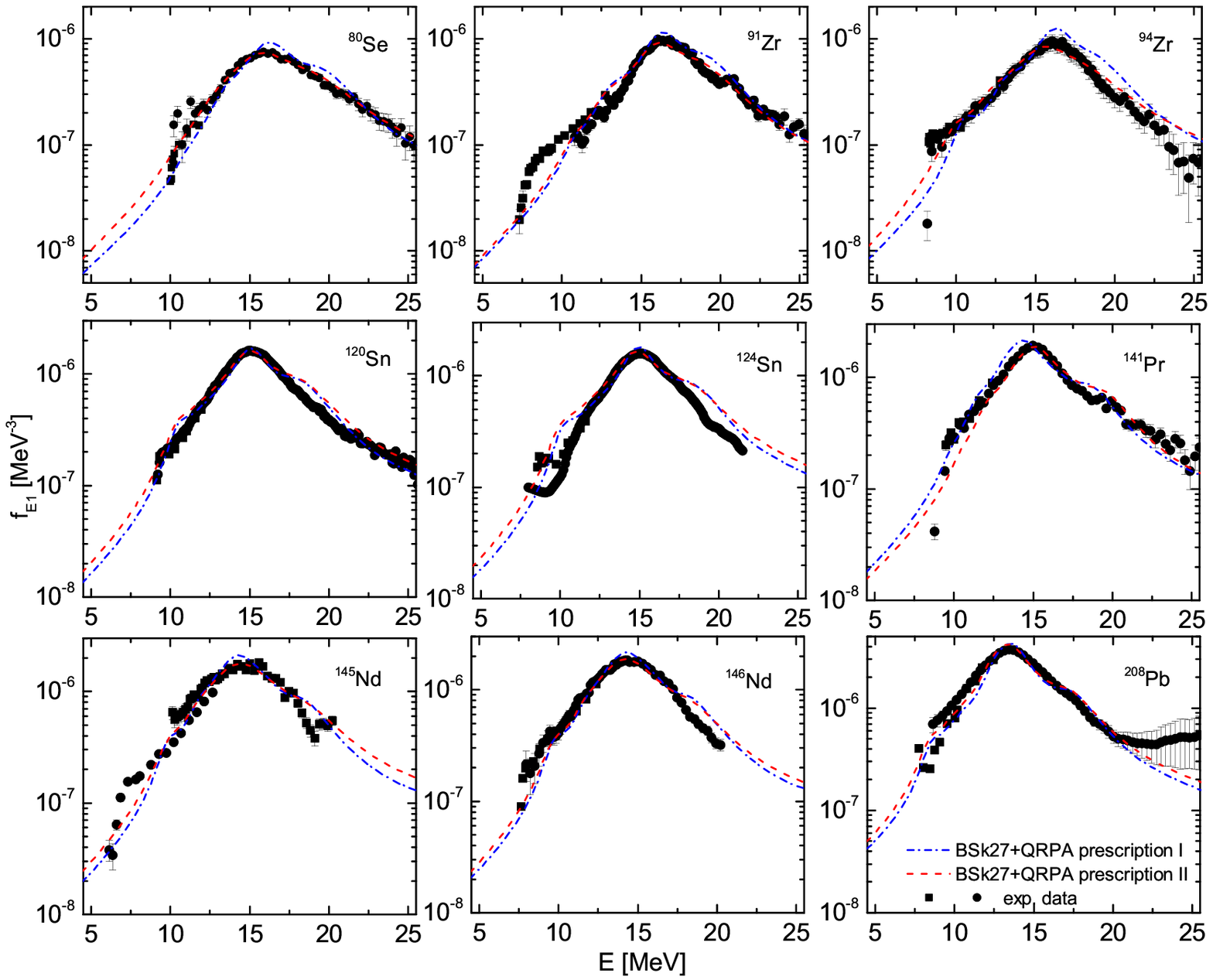}
\vskip -0.3cm
\caption{\label{fig4-absdata} Same as Fig.~\ref{fig2-abs} for another 9 spherical nuclei for which recent photoneutron emission cross sections have been estimated \cite{dataZr,dataSn,dataPr,dataNd,dataPb}.}
\end{figure*}

\section{Large-scale calculations of the E1 photon strength functions}
\label{calcompdis}

\subsection{Damping procedure for E1 photon strength function}
\label{sec_3a}

It is known that the giant dipole resonance (GDR) is experimentally characterized by a relatively large width and a corresponding finite lifetime. The QRPA calculations satisfactorily reproduce the GDR centroid energy and the fraction of the energy-weighted sum rule exhausted by the E1 mode. However, it is crucial to go beyond this approximation scheme to describe  properly the damping properties of the collective motion. Such state-of-the-art calculations including effects beyond the 1p-1h excitations and phonon coupling  \cite{Colo01,Sarchi04,Tsoneva08,Papakonstantinou09,Litvinova13,Achakovskiy15,Egorova16,Gambacurta18} are now available but  remain computer-wise intractable for large-scale applications, and extremely challenging for heavy deformed nuclei, in particular for systems with an odd number of nucleons. For this reason, a semi-empirical broadening of the GDR width has been applied to the present QRPA calculation.

The procedure used in the present work to damp the E1 photon strength function can be found in Refs.~\cite{damping,BSk7}. The E1 photon strength function $S_{E1}(E)$ (in $e^{2} {\rm fm}^{2} {\rm MeV}^{-1}$) is derived from the QRPA strength distribution $S(\omega)$ by folding it with a Lorentzian function $L(E,\omega)$, {\it i.e.}
\begin{equation}
S_{E1}(E) =\sum_{\sigma}L(E,\omega)S(\omega),
\label{e11}
\end{equation}
where $L(E,\omega)$ is the Lorentzian function given by
\begin{equation}
L(E,\omega)=\frac{1}{\pi}\frac{E^{2}\Gamma(E,C_{G})}{[E^{2}-(\omega-\Delta(E,C_{E}))^2]^{2}+\Gamma(E,C_{G})^{2}E^{2}},
\label{e12}
\end{equation}
in which $\Gamma(E,C_{G})$ is the width at half maximum and $\Delta(E,C_{E})$ allows for an energy shift of the centroid. The energy-dependent width $\Gamma(E,C_{G})$ can be calculated from the measured decay width of particle ($\gamma_{p}$) and hole ($\gamma_{h}$) states
\begin{equation}
\Gamma(E,C_{G})=\frac{1}{E}\int_{0}^{E}(1+C_{G})[\gamma_{p}(\epsilon)+\gamma_{h}(\epsilon-E)]d\epsilon,
\label{e13}
\end{equation}
and $\Delta(E,C_{E})$ can be obtained from $\Gamma(E,C_{G})$ by a dispersion relation \cite{damping}.

The interference factors $C_{E}$ and $C_{G}$ introduced in $\Gamma(E,C_{G})$ and $\Delta(E,C_{E})$, respectively, are due to the screening corrections of the exchange interaction which can interfere destructively with self-energy diagrams \cite{damping}. The microscopic evaluation of the interference factor is delicate. In practice, the values of $C_{E}$ and $C_{G}$ can be phenomenologically determined through a fit to experimental GDR data. Details of such a fitting procedure are given in the following subsection.

\subsection{Adjustment of the interference factors on experimental photodata}
\label{sec_3b}

The photo-induced reaction cross sections and the observed GDR parameters of peak energy, peak cross section and full width at half maximum, measured by bremsstrahlung, quasi-monoenergetic and tagged photons, have been compiled in Refs.~\cite{gdrdata1,gdrdata2,Plujko11,Plujko18} and provide the most relevant and reliable source for the determination of the interference factors. In practice, the E1 photon strength function $f_{E1}(E) [{\rm MeV}^{-3}] = S_{E1}(E) [{\rm mb/MeV}] \times 8.67373 \times 10^{-8}$ can be simply related to the photoabsorption cross section $\sigma_{abs}$ \cite{gdrdata1,gdrdata2,Plujko11,Plujko18} through
\begin{equation}
f_{E1} (E)=\frac{\sigma _{E1} (E)}{3E \left(\pi \hbar c\right)^{2} }.
\label{e14}
\end{equation}
Since the BSk27 $ph$ QRPA strength distribution $S(\omega)$ used to derive the strength function $S_{E1}(E)$ in Eq.~(\ref{e11}) is calculated in spherical symmetry, only experimental data of spherical nuclei are considered in the fitting procedure. For this reason, we restrict ourselves here to fit the theoretical results predicted by Eqs.~(\ref{e11}-\ref{e13}) to experimental photoabsorption strength functions for 48 spherical nuclei by adjusting the interference factors $C_{E}$ and $C_{G}$. Two prescriptions (so-called prescription I and prescription II) are proposed to extract $C_{E}$ and $C_{G}$.

In the first prescription (prescription I), a unique constant value $C_{cst}$ is adopted for both interference factors $C_{E}$ and $C_{G}$ for all nuclei. The value of $C_{E}$ = $C_{G}$ = $C_{cst}$ = -0.655 is obtained by minimizing the sum $S_{\Sigma}$ defined as
\begin{equation}
S_\Sigma =\sum_{i=1}^{N}(\sigma^{exp}(E_{i})-\sigma^{QRPA}(E_{i})){^2}.
\label{e15}
\end{equation}
In Eq.~(\ref{e15}), $\sigma^{exp}$ is the experimental photoabsorption cross sections and $\sigma^{QRPA}$ is the corresponding QRPA predictions. $E_{i}$ is picked up in sequence of the GDR energy range from $E_{min}$ to $E_{max}$ with the interval of 0.1 MeV, and $N=(E_{max}-E_{min})/(0.1 {\rm MeV})$. Only data within the energy range $[E_{min},E_{max}]$ defining the GDR region, as suggested by Refs.~\cite{gdrdata1,gdrdata2,Plujko11,Plujko18}, are taken into account.

In the second prescription (prescription II), another strategy is followed. In particular, different possible values of the interference factors $C_{E}$ and $C_{G}$ are considered for each nucleus. Fits to the experimental photon strength function from photodata are  conducted for the 48 spherical nuclei by minimizing the sum $S_\Sigma$ (Eq.~\ref{e15}) to derive individual values $C_{E}$ and $C_{G}$.
%The GDR energy ranges used for the individual fitting in prescription II are identical to those used for the global fitting in prescription I.

The resulting interference factors $C_{E}$ and $C_{G}$ are shown in Fig.~\ref{fig1-cecg} for the 48 spherical nuclei considered. The constant value $C_{E}$ = $C_{G}$ = -0.655 for prescription I is shown by the dash-dot line, while values for the $A$-dependent prescription II are depicted by triangles. The interference factors impact both the width and the centroid energy of the GDR. Within the present damping method, the peak energies around $E = 15$~MeV can be shifted by $5\times (1+C_{E})$~MeV upwards approximately. Moreover, if $C_{G} \geq -0.2$, the GDR broadening becomes too large and incompatible with experimental photoabsorption data, while if $C_{G} \leq -0.8$, the fine structure inherent to the 1p-1h QRPA estimate cannot be adequately smeared out and leads to photon strength functions again incompatible with experimental data.

\begin{figure*}
\hskip -0.6cm
\includegraphics[width=17cm]{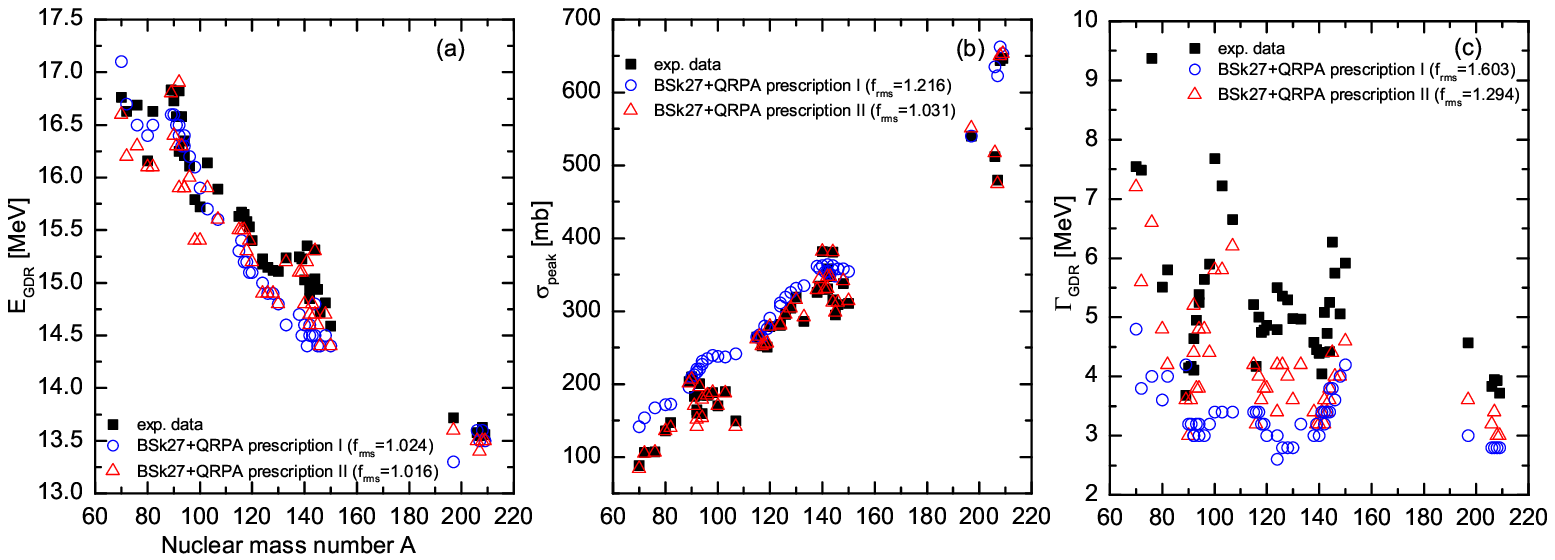}
\vskip -0.3cm
\caption{\label{fig5-GDRcomparison} Comparisons of the experimental GDR properties (black square) compiled in RIPL-3 database and the BSk27 QRPA results calculated by the interference factors of prescriptions I (blue circle) and II (red triangle) for the 48 spherical nuclei. The position of peak energy ($E_{GDR}$), the strength (represented by the photoabsorption cross section) at the peak energy ($\sigma_{peak}$), and the full width at half maximum ($\Gamma_{GDR}$) are respectively shown in panels (a), (b) and (c).}
\end{figure*}

\subsection{Comparison with experimental data}
\label{sec_3c}

Based on Eqs.~(\ref{e11}-\ref{e13}), 2 sets of photon strength functions for the 48 spherical nuclei are  calculated using the interference factors $C_{E}$ and $C_{G}$ obtained with prescriptions I and II, respectively. In Figs. \ref{fig2-abs}, \ref{fig3-abs} and \ref{fig4-absdata}, our 2 sets are compared with experimental photoabsorption data \cite{Goriely19,gdrdata1,gdrdata2,Plujko11,Plujko18}, as well as data extracted from recent photoneutron cross sections \cite{dataZr,dataSn,dataPr,dataNd,dataPb}. Our previous BSk7+QRPA predictions \cite{BSk7} are also shown for comparison. It can be seen that experimental GDR centroid energies and widths are overall rather well reproduced. However, as observed in Fig. \ref{fig2-abs}, some non-negligible discrepancy between the predictions and experimental data is visible for  light nuclei with mass number $A \la 100$. The description of such nuclei remains difficult in any global model, including the Lorentzian approach, and would require future improvements. Similarly to the BSk7+QRPA results, some extra strength located about 3-4~MeV above the GDR peak (sometime also experimentally observed) is systematically found in the present BSk27+QRPA calculations. Prescription II is also seen to lead to a better description of data in comparison with prescription I.

Furthermore, two sets of the GDR parameters, {\it i.e.} the position of the peak energy ($E_{GDR}$), the full width at half maximum (FWHM, $\Gamma_{GDR}$), and the strength (represented by the photoabsorption cross section) at the peak energy ($\sigma_{peak}$) have been extracted from the calculated BSk27+QRPA E1 photon strength functions with prescriptions I and II. These GDR properties are compared with experimental values \cite{RIPL3,Plujko18} in Fig.~\ref{fig5-GDRcomparison} for the 48 spherical nuclei. Deviations between theoretical and experimental data can be characterized by a root mean square (rms) factor $f_{rms}$ defined as usual as (in this case for the peak cross section $\sigma_{peak}$)
\begin{eqnarray}
f_{rms}=\exp\Big[\frac{1}{N_{exp}}\sum_{i=1}^{N_{exp}}\big(\ln\frac{\sigma_{peak}^{QRPA}(i)}{\sigma_{peak}^{exp}(i)}\big)^{2}\Big]^{1/2},
\label{e16}
\end{eqnarray}
where $N_{exp} = 48$ is the number of spherical nuclei included in our study. Similar $f_{rms}$ expression can be derived for $E_{GDR}$ and $\Gamma_{GDR}$ by simply replacing $\sigma_{peak}$ with $E_{GDR}$ and $\Gamma_{GDR}$ in Eq.~(\ref{e16}). The resulting $f_{rms}$ values are given in Fig.~\ref{fig5-GDRcomparison} for each GDR parameter and for both prescriptions I and II.

For the peak position $E_{GDR}$ in Fig.~\ref{fig5-GDRcomparison}(a), we obtain an rms deviation $f_{rms}$=1.024 with prescription I and 1.016 for prescription II. This means that the GDR centroid are globally predicted within 2\%. Fig.~\ref{fig5-GDRcomparison}(b) shows that the GDR experimental peak cross section are determined with an rms deviation $f_{rms}=1.261$ for prescription I and 1.031 for prescription II. Obviously, a better accuracy is obtained with prescription II since in this case the interference factors are adjusted individually for each nucleus, but globally both prescriptions reproduce rather well the experimental trend. In contrast, more discrepancies can be seen (Fig.~\ref{fig5-GDRcomparison}c) between the GDR theoretical and experimental FWHM. Here also prescription II gives rise to a significantly better description of the experimental FWHM with an $f_{rms}\simeq 1.3$ with prescription II compared to 1.6 for prescription I.

\begin{figure*}
\hskip -0.6cm
\includegraphics[width=17cm]{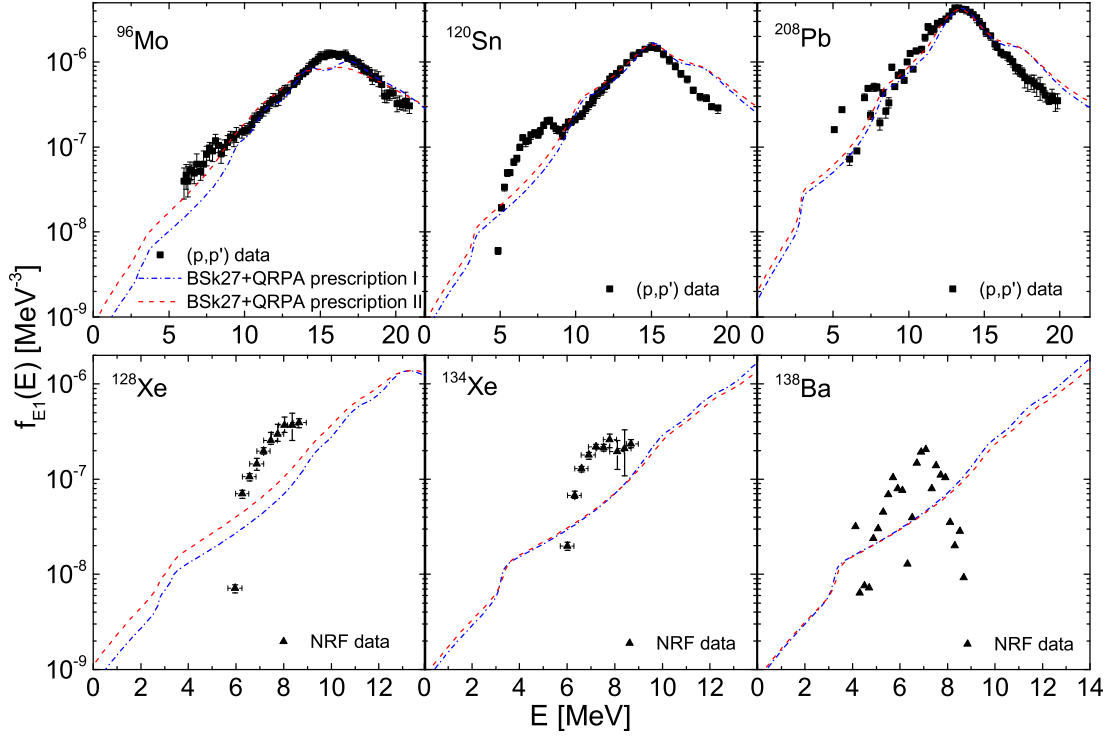}
\vskip -0.3cm
\caption{\label{fig6-NRFpp} Comparisons of  proton scattering  \cite{ppMo96,ppSn120,ppPb208} and NRF \cite{NRFXe,NRFBa138} data with BSk27+QRPA results obtained with prescriptions I and II.}
\end{figure*}

In Fig.~\ref{fig6-NRFpp}, the BSk27+QRPA photon strength functions are compared with  experimental E1 data at  low energies available from   nuclear resonance fluorescence (NRF) \cite{NRFXe,NRFBa138}  and proton scattering \cite{ppMo96,ppSn120,ppPb208} experiments. A relatively good agreement between theory and experiment can be seen in Fig.~\ref{fig6-NRFpp}. Deviation as well as uncertainties associated with NRF experiments are discussed in Ref.~\cite{Goriely19}.

Finally, note that most of the other experimental data available to test prediction of the photon strength function \cite{Goriely19} are also sensitive to the M1 photon strength function and can therefore not be used directly in the present work to learn more about the predictive power of our BSk27+QRPA model. The future estimate of the M1 photon strength function within the same framework and with the same interaction will allow us to include such experimental data for comparison.

\subsection{Interference factors for experimentally unknown nuclei}
\label{sec_3d}

A major question arises when having to assign the interference factors to the many experimentally unknown nuclei. In the case of prescription I, it seems obvious that a constant value $C_{E}= C_{G} = -0.655$ should be adopted for all the $\sim 10000$ nuclei of interest. For prescription II, in contrast and as shown in Fig.~\ref{fig1-cecg}, the interference factors are clearly nucleus dependent without any evident pattern as a function of the neutron number ($N$) or atomic mass number ($A$). Consequently, extracting globally a smooth $N$ or $A$ dependence would inevitably give rise to significant discrepancies if applied to known nuclei. For this reason, it is proposed to carry an interpolation to deduce the interference factors for nuclei for which no GDR data is available on the basis of the known interference factors for the 48 spherical nuclei. The GDR parameters,  in particular the FWHM, are known to be shell dependent and are essentially related to the neutron number \cite{Gogny2016}, hence the interpolation is performed as a function of the neutron number $N$. Note that the interference factors of the two nuclei with the minimum and maximum neutron numbers among the 48 spherical nuclei (here $N_{min}=38$ and $N_{max}=126$) are respectively used for nuclei with neutron numbers smaller than $N_{min}$ or larger than $N_{max}$.

\subsection{Nuclear deformation effect}
\label{sec_3e}

For  deformed nuclei, the GDR is known to split into two modes as a result of the different resonance conditions characterizing the oscillations of protons against neutrons along the axis of rotational symmetry and an axis perpendicular to it. In the phenomenological approach, the Lorentzian-type formula is generalized to a sum of two Lorentzian-type functions of peak energies $E_{GDR}^{l}$ and widths $\Gamma_{GDR}^{l}$ \cite{BSk7,Plujko18,RIPL3}. The peak energies $E_{GDR}^{l}$ ($l=1,2$) satisfy the relations
\begin{eqnarray}
E_{GDR}^{1}+2E_{GDR}^{2}=3E_{GDR},\nonumber\\
E_{GDR}^{2}/E_{GDR}^{1}=0.911\eta+0.089,
\label{e17}
\end{eqnarray}
and each width $\Gamma_{GDR}^{l}$ ($l$=1,2) is given the same dependence as the respective peak energy $E_{GDR}^{l}$. In Eq.~(\ref{e17}), $\eta$ is the ratio of the diameter along the axis of symmetry to the diameter along an axis perpendicular to it.
%Experimentally, it is also found that  the high-energy peak at $E_{GDR}^{2}$ is characterized by twice the strength of the low-energy one at $E_{GDR}^{1}$.
%Furthermore, the strength around the low-lying energy can be spread because of the splitting of GDR.

\begin{figure*}
\hskip -0.6cm
\includegraphics[width=17cm]{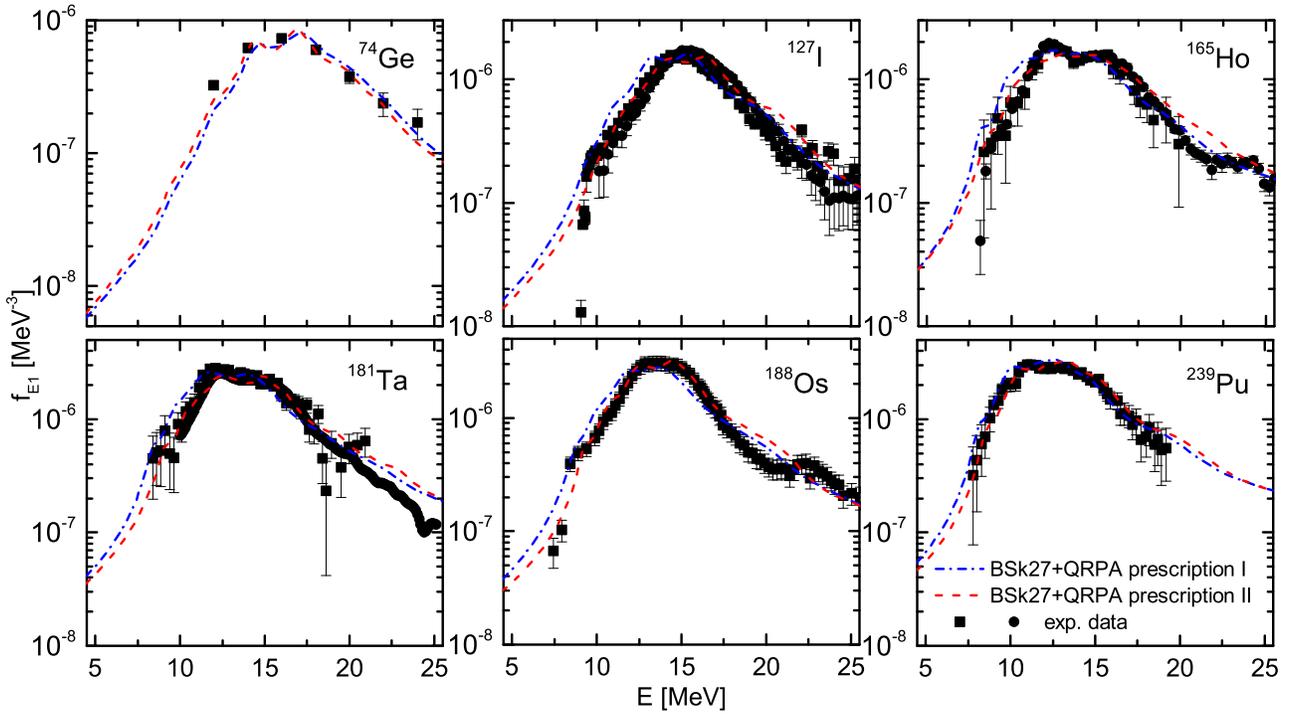}
\vskip -0.3cm
\caption{\label{fig7-absdefdata} Comparison between experimental \cite{Goriely19} and BSk27+QRPA  photon strength functions for 6 deformed nuclei. Both prescriptions I (dash-dot line) and II (dashed line) are shown.}
\end{figure*}

In the present study, the deformation effect is implemented into the damping procedure of Eqs.~(\ref{e11}) and (\ref{e12}), in which the Lorentzian function at a given energy $E$ is split into two Lorentzian functions centered according to Eq.~(\ref{e17}) and characterized by a width $\Gamma(E)$ (see Eq.~\ref{e13}) obtained from the same relations as Eq.~(\ref{e17}). Note that deformation parameters $\eta$ from the HFB-27 mass model \cite{hfb27} are consistently considered and that only nuclei with a quadrupole deformation parameter $|\beta_{2}| \geq 0.15$ are regarded as deformed. The resulting photon strength functions are  compared with experimental data \cite{gdrdata2,dataGe1,dataGe2,dataTa1,dataTa2,dataTa3,dataOs} for 6 deformed nuclei in Fig.~\ref{fig7-absdefdata}. Globally, the experimental splitting are rather well reproduced, especially for the well deformed nuclei $^{165}$Ho and $^{181}$Ta.
%The larger distribution of the GDR strength at the higher peak energy is reflected by the present calculation, at least in the rare-earth region.

\begin{figure}
\hskip -0.6cm
\includegraphics[width=9cm]{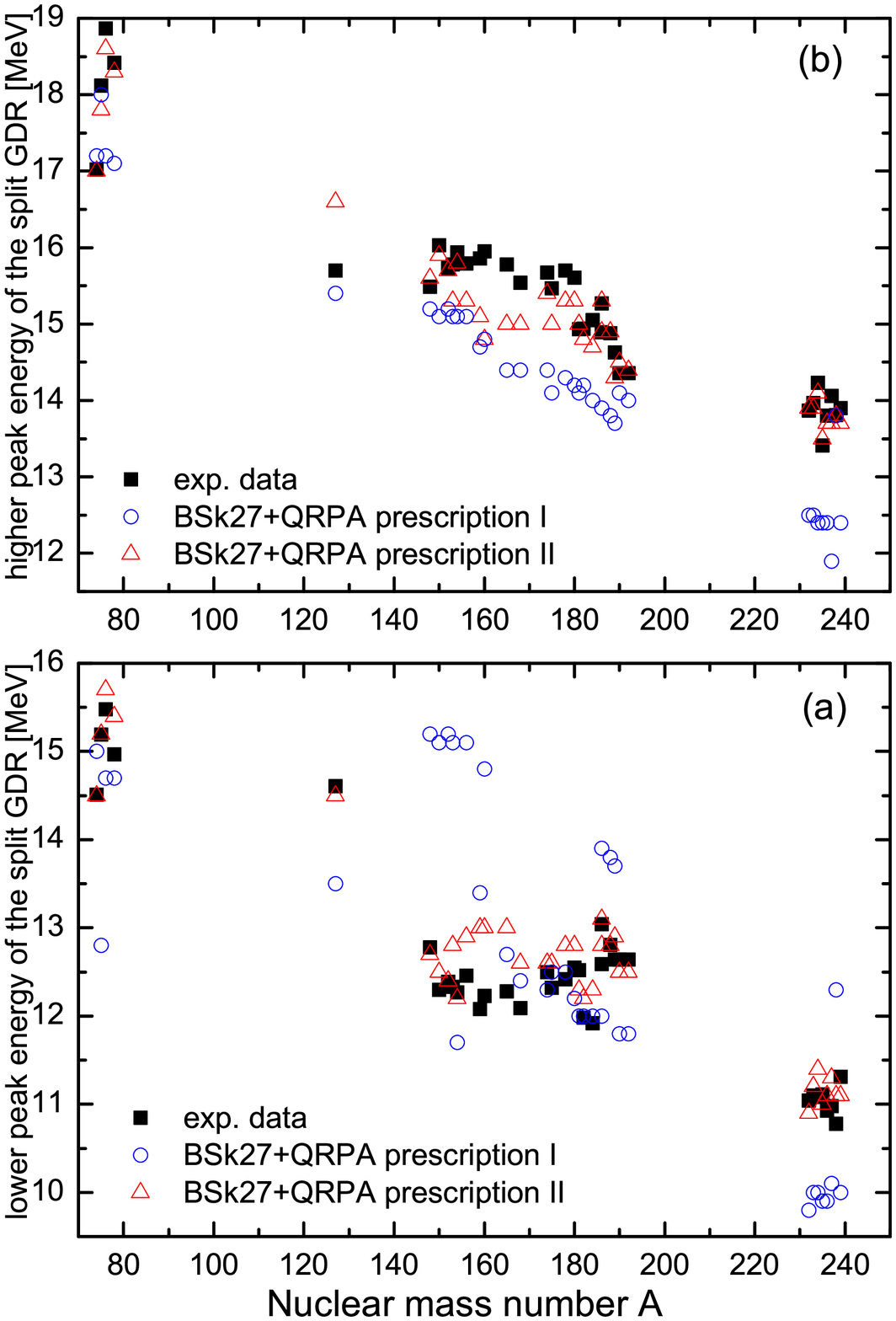}
\vskip -0.3cm
\caption{\label{fig8-2peaks} Comparison between experimental and BSk27+QRPA peak energies for 36 deformed nuclei. Both prescriptions I (blue circles) and II (red traingles) are shown. The low- and high-energy peak energies are shown in (a) and (b), respectively.}
\end{figure}

To test the predictive power of the present approach for deformed nuclei, we compare in Fig.~\ref{fig8-2peaks} the centroid energies of the two peaks predicted by BSk27+QRPA with known GDR parameters \cite{RIPL3,Plujko18} for 36 deformed nuclei. Here also, globally the peak energies can be rather well described, in particular by our prescription II, prescription I usually overestimating the energy of the first peak for rare-earth nuclei and underestimating it for actinides.
For this reason,  the subsequent calculations will be carried out based on the interference factors of prescription II that overall leads to significantly better descriptions of the GDR experimental features.

\subsection{Temperature-dependent correction and low-energy strength}
\label{sec_3f}

The BSk27+QRPA results obtained up to now focused on the photoabsorption at energies characteristic of the GDR region.
However, some experimental studies \cite{Brink1,Brink2,Brink3} indicate that the photon strength function may also depend on the excitation energy of the initially decaying state and consequently that the de-excitation (downwards) strength function may differ from the photoabsorption (upwards) one. This violation of the so-called Brink hypothesis can be described by introducing a temperature-dependent correction to the photoabsorption strength function (see {\it e.g.} \cite{yadfiz1983,Kopecky1990,BSk7}).

%knowledge of the low-energy tail of the photon strength function, especially around the neutron separation energy, is crucial to describe the photonuclear and capture reactions (especially the neutron capture). At the low energies, experimental E1 strength is obtained from the $\gamma$-decay data \cite{decaydata1,decaydata2} that are connected with the "downwards" E1 photon strength function, and the temperature-dependent correction traditionally introduced into the phenomenological GDR width \cite{Kopecky1990} usually deals with the $\gamma$-decay data. On the other hand, the $\gamma$-absorption data used for the present study so far are related to the "upwards" E1 photon strength function, so the compatibility of the temperature-dependent correction with the $\gamma$-absorption data should be satisfied.

In the present study, two temperature-dependent corrections (Correction I and Correction II) as inspired from the Fermi liquid theory \cite{yadfiz1983} are considered. For both corrections, a temperature-dependent term is explicitly added to the energy-dependent width $\Gamma(E)$ (see Eq.~\ref{e13}), namely
\begin{eqnarray}
\Gamma^\prime(E,T)=\Gamma(E)+\frac{4\pi\alpha T^{2}\Gamma(E)}{E_{GDR}E} \nonumber\\
\label{e18}
\end{eqnarray}
and
\begin{eqnarray}
\Gamma^\prime(E,T)=\Gamma(E)+\frac{4\pi^{2}T^{2}\Gamma_{GDR}}{E_{GDR}^{2}} \nonumber\\
\label{e19}
\end{eqnarray}
for Correction I and Correction II, respectively. In Eqs.~(\ref{e18}) and (\ref{e19}), $T$ refers to the nuclear temperature of the absorbing state, and $\alpha$ is a normalization constant. In Eqs.~(\ref{e18}-\ref{e19}), for simplicity, mean values of the GDR peak energy and width for the 48 spherical nuclei are adopted, namely $E_{GDR}$=15.5MeV and $\Gamma_{GDR}$=5.2 MeV and the normalization constant $\alpha$ is set to 2 \cite{BSk7}. The modified width $\Gamma'(E,T)$ is imported into the damping procedure to determine the de-excitation E1 photon strength function, guaranteeing the compatibility with $\gamma$-absorption data.

Taking into account both expressions of the temperature-dependent corrections for the width $\Gamma^\prime(E,T)$ given by Eq.~(\ref{e18}) (Correction I) and Eq.~(\ref{e19}) (Correction II), two sets of temperature-dependent BSk27+QRPA E1 photon strength functions have been computed for temperatures ranging between $T=0$ ({\it i.e.} without correction) to $T=2.0$~MeV. In such calculations, both the interference factors of prescription II (see Sec.~\ref{sec_3d}) and the deformation effect (see Sec.~\ref{sec_3e}) are simultaneously considered.

\begin{figure*}
%\hskip -0.6cm
\includegraphics[width=17cm]{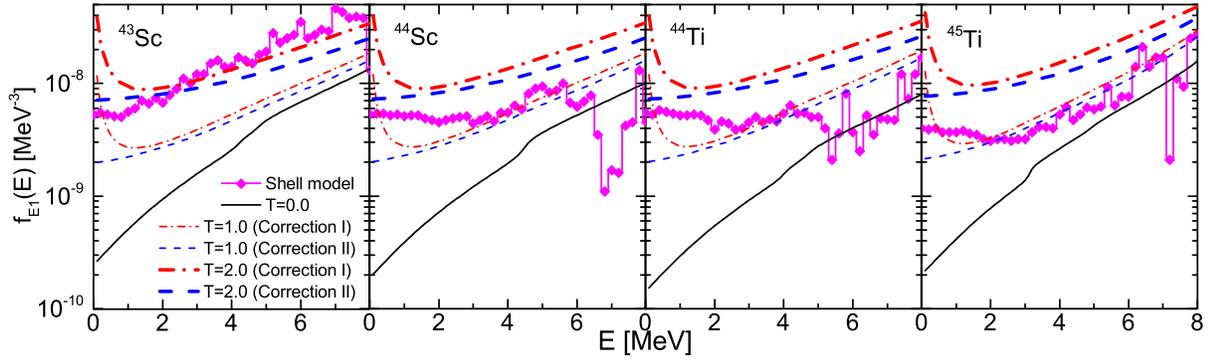}
%\vskip -0.3cm
\caption{\label{fig9-tem} Comparison between the  shell model E1 strength function \cite{SM,Sieja18} and the temperature-dependent BSk27+QRPA results obtained with corrections I and II at $T=0$, 1.0 and 2.0~MeV for 2 isotopes of Sc and 2 isotopes of Ti.}
\end{figure*}

To test the temperature dependence, we compare in Fig.~\ref{fig9-tem} both corrections at different temperatures (namely $T=1$ and 2~MeV) with the theoretical results obtained by the shell model \cite{SM,Sieja18} for 4 light nuclei $^{43}$Sc, $^{44}$Sc, $^{44}$Ti, and $^{45}$Ti. It can be seen  in Fig.~\ref{fig9-tem} that the shell model calculations are fairly well reproduced by the temperature-dependent BSk27+QRPA results.
The E1 photon strength functions below the neutron separation energies are dramatically affected by the temperature correction. A non-zero behavior with a significant increase and broadening of the E1 strength function is obtained for  $E \rightarrow 0$. It can also be seen that in the case of Correction I, the extra enhancement of the E1 strength function resulting from the $1/E$ dependence of Eq.~(\ref{e18}) can be observed as an ``upbend'' behavior at $E\rightarrow 0$. Note that the upbend of the strength function observed experimentally \cite{Voinov04,Guttormsen05} has  been assumed to be of the $M1$ nature following  shell-model predictions \cite{Schwengner13}, though no experimental evidence exists for the moment.
However, the ``upbend'' behaviour of the temperature-dependent E1 strength at E $\rightarrow$ 0 obtained with Eq.~(\ref{e18}) (Correction I) is clearly not confirmed by the shell model. For this reason, in the study to follow, the temperature-dependent correction II is adopted.

\begin{figure}
\hskip -0.6cm
\includegraphics[width=9cm]{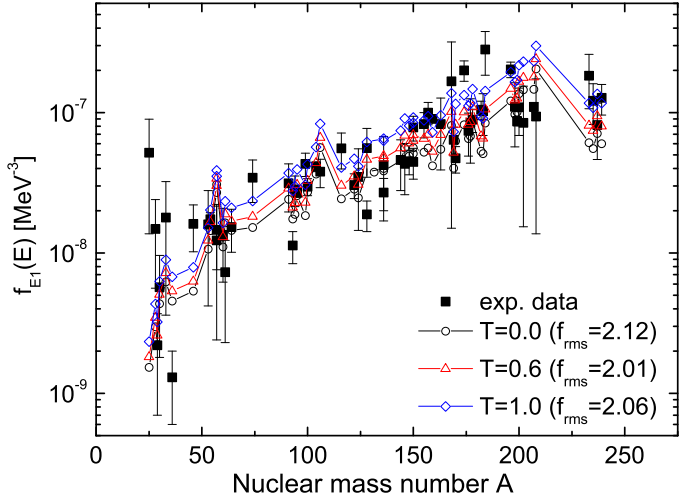}
\vskip -0.3cm
\caption{\label{fig10-temsys} Systematic comparison between the experimental E1 strength $f_{E1}$ at the energies ranging from 4 to 8 MeV in RIPL3 compilation  for 60 nuclei from $^{25}$Mg to $^{239}$U \cite{RIPL3} and the BSk27+QRPA results calculated by Eq.~(\ref{e19}) (Correction II) at different temperatures of $T=0$, 0.6 and 1.0 MeV.}
\end{figure}

\begin{figure}
\hskip -0.6cm
\includegraphics[width=9cm]{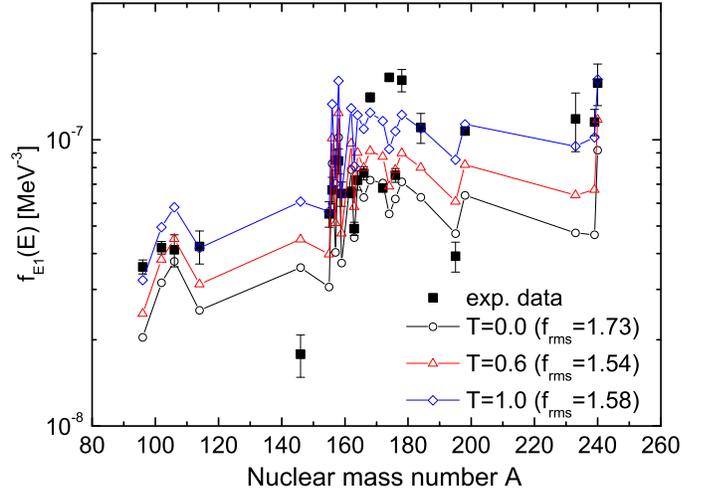}
\vskip -0.3cm
\caption{\label{fig11-temsys} Systematic comparison between ARC E1 strength function for 25 nuclei from $^{96}$Mo to $^{240}$Pu at the energies ranging from 4 to 8 MeV \cite{ARC} and BSk27+QRPA results calculated by Eq.~(\ref{e19}) (Correction II) at 3 different temperatures of $T=0.0$, 0.6 and 1.0 MeV.}
\end{figure}

At the low energies ranging from 4 MeV to 8 MeV, the BSk27+QRPA E1 strength $f_{E1}$ calculated by Eq.~(\ref{e19}) (Correction II) with different temperatures of $T=0.0$, 0.6 and 1.0 MeV are systematically compared with the compiled experimental data for 60 nuclei from $^{25}$Mg to $^{239}$U \cite{RIPL3} in Fig.~\ref{fig10-temsys} and for 25 nuclei from $^{96}$Mo to $^{240}$Pu \cite{ARC} in Fig.~\ref{fig11-temsys}, respectively. The data sets used for the comparison include resolved-resonance measurements, thermal captures measurements and photonuclear data. In some cases the original experimental data need to be corrected, typically for non-statistical effects, so that only the results recommended by Refs. \cite{RIPL3,ARC,Kopecky17} are considered in Figs.~\ref{fig10-temsys} and \ref{fig11-temsys}. We find that all the BSk27+QRPA results with and without temperature-dependent correction reproduce well the trend of the experimental E1 strength within the error bars. In particular, the rms deviation $f_{rms}$ (see Eq.~\ref{e16}) on the ratios of the theoretical predictions to the experimental data are estimated to $f_{rms}$= 2.12, 2.01, and 2.06 for $T=0.0$, 0.6, and 1.0~MeV, respectively, for the 60 nuclei from $^{25}$Mg to $^{239}$U compiled in RIPL3 \cite{RIPL3}, and $f_{rms}$ = 1.73, 1.54, and 1.58 for the 25 nuclei from $^{96}$Mo to $^{240}$Pu from  the recent analysis of the average resonance capture (ARC) measurements \cite{ARC,Kopecky17}. Such a degree of accuracy is similar to the one obtained with the previous BSk7+QRPA calculation \cite{BSk7} and the Gogny D1M+QRPA calculation \cite{Gogny2016}. The $f_{rms}$ values indicate that the BSk27+QRPA calculation taking into account the temperature-dependent correction is globally in better agreement with the $\gamma$-decay data, though the calculation without temperature correction remains compatible with  experimental results.

\subsection{Comparisons with other predictions of E1 strength}

For all the $\sim10000$ nuclei with $8 \leq Z \leq 124$ lying between the proton and the neutron drip lines, we have calculated a complete set of  temperature-dependent BSk27+QRPA E1 photon strength functions. These have been obtained through the damping procedure of Eqs.~(\ref{e11}-\ref{e13}) (Sec.~\ref{sec_3a}) with the interference factors of prescription II (see Sec.~\ref{sec_3b}-\ref{sec_3d}), the deformation effects (Sec.~\ref{sec_3e}), as well as the temperature-dependent correction II (Sec.~\ref{sec_3f}).

\begin{figure*}
\hskip -0.6cm
\includegraphics[width=13cm]{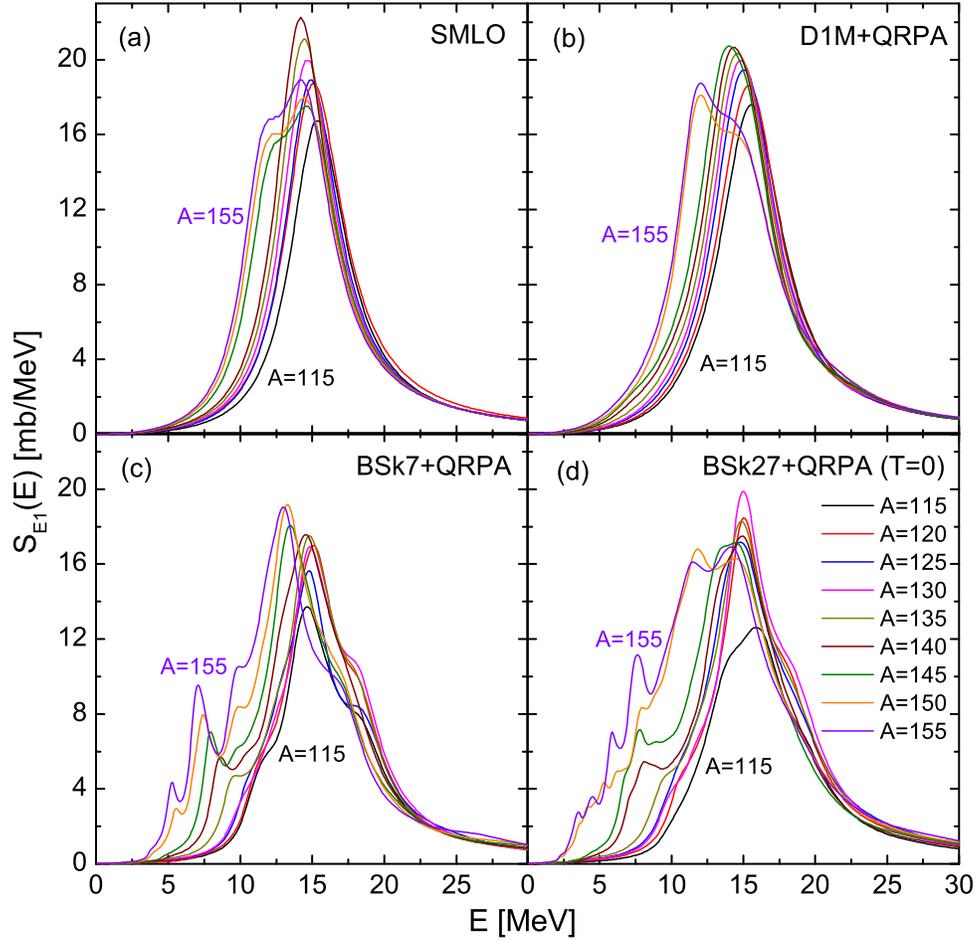}
\vskip -0.3cm
\caption{\label{fig12-strcomp} The E1 photon strength functions for the isotope sof Sn (from A=115 to A=155 with a step of $\Delta$A=5) obtained from (a) the empirical Lorentzian model SMLO \cite{Plujko18,Goriely19b}, (b) the D1M+QRPA \cite{Gogny2016,Goriely19}, (c) the BSk7+QRPA \cite{BSk7}  and (d) the present BSk27+QRPA.}
\end{figure*}

We compare in Fig.~\ref{fig12-strcomp} our BSk27+QRPA E1 photon strength function in the isotopic chain of Sn (from $A=115$ to $A=155$ with a step of $\Delta A=5$) with the empirical simple modified Lorentzian (SMLO)  \cite{Plujko18,Goriely19b},  D1M+QRPA  \cite{Gogny2016,Goriely19} and BSk7+QRPA \cite{BSk7}  models. Note that the SMLO model \cite{Plujko18,Goriely19b} is an updated version of the generalized Lorentzian model  that has been thoroughly compared with all experimental data available. Our new BSk27+QRPA results are rather similar to the other QRPA predictions and the empirical Lorentzian approximation for nuclei close to the valley of $\beta$-stability. However, in the neutron-rich region, the present BSk27+QRPA calculations start deviating from the Lorentzian curves in the same way as the BSk7+QRPA used to. For the Sn isotopes above $N = 82$ shell closure, in particular, some extra strength is found to be located in the low-energy range of [5,10]~MeV. The more exotic the nucleus, the stronger this low-energy E1 strength.
%Such extra strength in the low-energy range is also predicted by the calculations of BSk7-QRPA \cite{BSk7} and D1M-Gogny QRPA\cite{Gogny2016}, but not reproduced by the empirical Lorentzian approach \cite{RIPL3} for the exotic neutron-rich nuclei.

\section{Calculations of the astrophysical neutron capture reaction rates}

The E1 photon strength function directly impacts the neutron capture reaction rate of astrophysical interest. To quantitatively investigate such an impact, systematical comparisons of the neutron capture reaction rates calculated with different E1 photon strength functions are performed for all the $\sim 10000$ nuclei with $8 \leq Z \leq 124$ lying between the proton and the neutron drip-lines. The reaction rates are estimated on the basis of the Hauser-Feshbach statistical model with the TALYS reaction code \cite{TALYS1,TALYS2,TALYS3}.
%TALYS provides a complete description of all reaction channels and observables, in which the reaction mechanisms of direct, pre-equilibrium and compound are simultaneously included, and many state-of-the-art nuclear models covering these reaction mechanisms are implemented.

For the nuclear ingredients used in TALYS, all experimental information are considered whenever available, and if not, various local and global microscopic (or semi-microscopic) nuclear models have been incorporated to represent the nuclear structure and interaction properties. Such a combination of experimental data and model predictions allows not only for the essential coherence of the predictions for all experimental unknown data, but also a rather reliable extrapolation away from experimentally known energy or mass regions, as required in specific applications like nuclear astrophysics. More specifically, the  experimental inputs of the nuclear masses \cite{Wang17}, the discrete energy levels \cite{RIPL3}, as well as the model predictions including the Skyrme-HFB nuclear masses \cite{hfb27}, the JLMB optical model potentials \cite{JLMB1,JLMB2} and the HFB+combinatorial nuclear level densities \cite{NLD} are considered. Furthermore, in order to keep the consistency, all these nuclear ingredients are identically used for calculations of the neutron capture rates with different E1 photon strength functions. Note, however, to study the impact of the E1 strength function only, the M1 contribution to the de-excitation mode is omitted in all calculations here.

\begin{figure}
\hskip -0.6cm
\includegraphics[width=9cm]{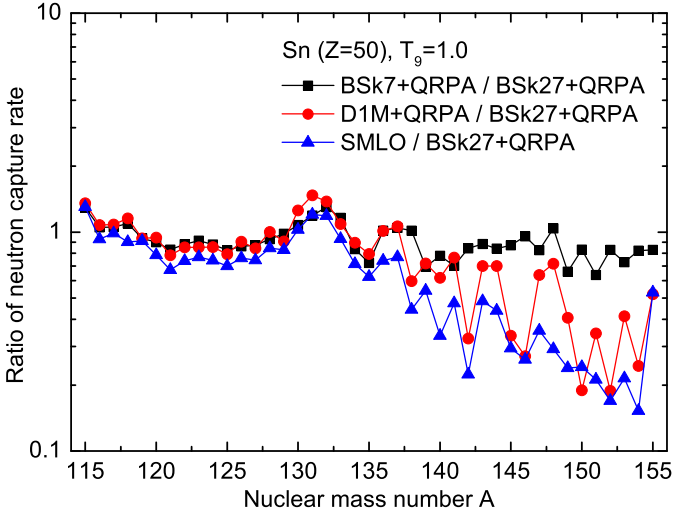}
\vskip -0.3cm
\caption{\label{fig13-Snrate} Ratios of the neutron capture reaction rates at the temperature of $T_{9}= 1$ calculated with the E1 BSk7+QRPA, D1M+QRPA and SMLO models to the one obtained with the present BSk27+QRPA model for the Sn isotopes in the mass range from $A=115$ to $A=155$.}
\end{figure}

Figure \ref{fig13-Snrate} compares the neutron capture reaction rates at $T_{9}=1$ (where $T_{9}$ is the temperature in unit of $10^{9}$ K) computed with the present BSk27+QRPA E1 photon strength function to those obtained with the BSk7+QRPA, D1M+QRPA and SMLO models for the Sn isotopes in the mass range from $A=115$ to $A=155$. We can see from Fig.~\ref{fig13-Snrate} that predictions with BSk27+QRPA  are very close to those obtained with BSk7+QRPA along the whole Sn isotopic chain. In contrast, BSk27+QRPA rates may be larger than D1M+QRPA or SMLO rates by a factor of about 5 for the most exotic neutron-rich Sn isotopes.

\begin{figure*}
\hskip -0.6cm
\includegraphics[width=15cm]{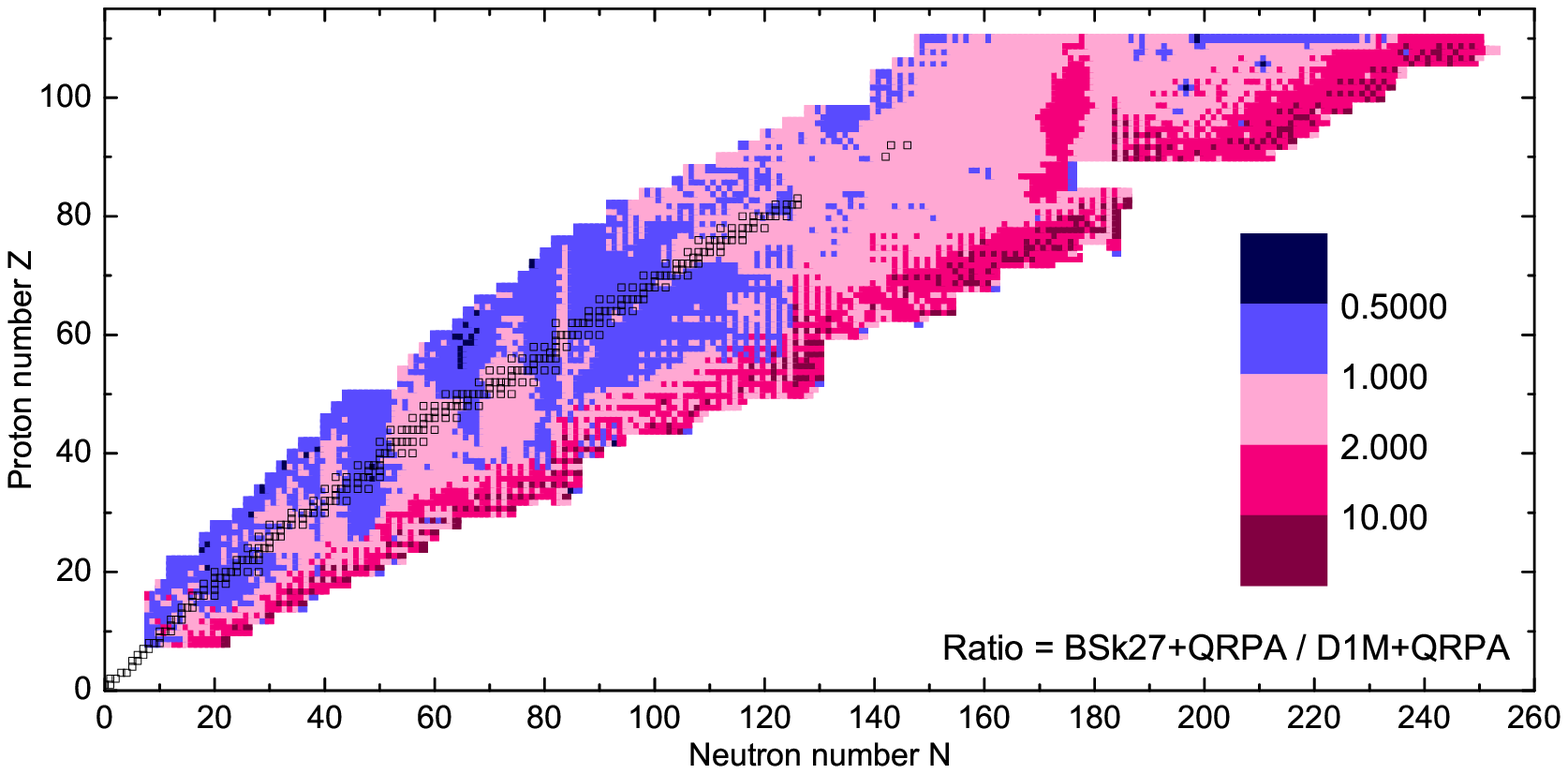}
\vskip -0.3cm
\caption{\label{fig14-allrate} Representation in the ($N,Z$) plane of the ratios of the neutron capture reaction rates at $T = 10^9$ K obtained with the present BSk27+QRPA E1 photon strength function to those obtained with the previous D1M+QRPA model. The open black squares correspond to the stable nuclei and very long-lived actinides.}
\end{figure*}

Similar deviations between the rates with BSk27+QRPA and D1M+QRPA can be observed for the other isotopic chains in Fig.~\ref{fig14-allrate} where rates are compared for all nuclei with $8 \leq Z \leq 110$ in the ($N,Z$) plane. Ratios between both predictions remain essentially within a factor of 2, except for the most neutron-rich nuclei where a factor of 10 at most can be reached, BSk27+QRPA giving rise to larger neutron capture rates with respect to D1M+QRPA.

\section{Summary}
\label{sect_sum}
Nuclear dipole excitations in the whole chart of nuclei are key ingredients for different nuclear applications, in particular for nuclear astrophysics. In the present study, the E1 photon strength functions for about 10000 nuclei with $8 \leq Z \leq 124$ lying between the proton and the neutron drip lines are systematically investigated combining simultaneously the microscopic HFB+QRPA model and the constraints from various available experimental information sensitive to the photon strength function.

In such an approach, the nuclear ground state is described through the BSk27 Skyrme effective nucleon-nucleon interaction which has been shown to provide so far the most accurate mass (and radii) predictions from mean-field models and at the same time to reproduce nuclear matter properties as estimated from realistic microscopic calculations \cite{hfb27}. In the QRPA framework, the nuclear excitation is considered as the collective superposition of two qp states built on top of the BSk27 HFB ground state, and the response function for the nuclear excitation, represented as the $ph$ strength distribution, is derived from the Green's function formalism. Large-scale calculations of the BSk27 QRPA $ph$ strength distributions for all the $\sim 10000$ nuclei with $8 \leq Z \leq 124$ have been performed assuming spherical symmetry.

Despite the predictive power of the HFB+QRPA approach, a number of phenomenological corrections needs to be included in order to describe effects beyond the standard 1p-1h HFB+QRPA model, but also to reproduce accurately available experimental data.
By folding the $ph$ strength distributions by a Lorentzian-type function, neglected effects, such as phonon coupling or 2p-2h excitations can be phenomenologically included. Such a folding procedure introduces two phenomenological interference factors, $C_{E}$ and $C_{G}$ that have been adjusted to reproduce at best experimental GDR parameters for spherical nuclei. Two prescriptions are proposed and tested, although only the most accurate one with an individual adjustment nucleus by nucleus has been retained. For nuclei for which no experimental data is available, interpolations of $C_{E}$ and $C_{G}$ between experimentally constrained values are performed as a function of the neutron number $N$.

%Two sets of the BSk27+QRPA E1 photon strength functions for all about 10000 nuclei with 8 $\leq$ Z $\leq$ 124 are computed based on the folding procedure with the interference factors of set I and set II, respectively. The GDR properties (e.g. the photoabsorption cross sections and the GDR properties of the peak energy and width) derived from the two sets of theoretical calculations are systematical compared with the available experimental observables. It is found that both sets of predictions agree fairly well with the experimental results, but with respect to the interference factors of set I, the interference factors of set II globally improve the agreement between the theoretical predictions and the experimental results.

To break the spherical symmetry, an empirical expression accounting for deformation effects is applied in the above-mentioned folding procedure of the QRPA photon strength function. The systematical comparison of the GDR parameter for 36 deformed nuclei show that deformation effects can be fairly well introduced by a simple split of the GDR structure, as classically done in empirical studies and also observed experimentally in photonuclear data.

Finally, a temperature-dependent correction on the width of folding Lorentzian function is introduced to take into account the possible non-zero energy of the initial absorbing state. Such a correction is fundamental to extract the de-excitation photon strength function from the photoabsorption strength calculated on the basis of the HFB+QRPA approach. Two specific corrections are introduced and tested against shell-model predictions and available low-energy data. Such temperature corrections lead to a significant increase and broadening of the E1 photon strength function below the neutron separation energies and is needed to reproduce experimental data derived from de-excitation channels, such as  ARC data.

Taking into all those corrections beyond the present HFB+QRPA approach, ({\it i.e.} effects beyond QRPA, deformation and temperature effects), a complete set of  E1 photon strength functions for about 10000 nuclei with $8 \leq Z \leq 124$ lying between the proton and the neutron drip lines are generated on the basis of the BSk27 interaction. This model has been shown to reproduce experimental photodata relatively well, at least for $A \ga 100$ nuclei, with an accuracy similar to what would have been obtained with a Lorentz-type approach globally fitted to the data \cite{Plujko18}. The E1 description for light $A \la 100$ nuclei remains difficult in any of the models available nowadays. Compared to the empirical Lorentzian formula, a systematic increase of the E1 strength function is found for neutron-rich nuclei. While a global agreement is found with respect to our previous BSk7+QRPA predictions, more strength is found at low energies relative to the axially deformed D1M+QRPA predictions. Such an extra low-energy strength has a direct impact on neutron capture cross sections and rates, the latter being found to be larger by up to a factor of 10 for exotic neutron-rich nuclei lying close to the neutron drip line.

Further improvements for the study of the photon strength function may be envisioned. In particular, in order to better describe the low-energy region, it is necessary to consider the magnetic dipole (M1) contribution. A complete E1+M1 photon strength functions predicted fully consistently within the same approach can allow us to broaden the comparison with available experimental data and to improve the estimate of radiative neutron capture and photoneutron rates of astrophysical interest. Experimental efforts to  further improve the description of the photon strength function, in particular with the future measurement of the photonuclear excitation at the Extreme Light Infrastructure - Nuclear Physics (ELI-NP), are promising \cite{ELINP}. It is believed that working along such a path is a way to further improve the study of r- and p-processes of nucleosynthesis on the basis of reliable and accurate nuclear physics inputs.

\begin{acknowledgements}
%This work is performed within the IAEA CRP on "Updating the Photonuclear Data Library and generating a Reference Database for Photon Strength Functions" (Project No. F410 32).
This work is carried out under the contract PN 19 06 01 05 sponsored by the Romanian Ministry of Research and Innovation. Y.X. acknowledges the supports from the ELI-RO program funded by the Institute of Atomic Physics (Magurele, Romania) under the contract ELI$\_$15/16.10.2020, and from the Extreme Light Infrastructure - Nuclear Physics (ELI-NP) - Phase II, a project co-financed by the Romanian Government and the European Union through the European Regional Development Fund - the Competitiveness Operational Programme (1/07.07.2016, COP, ID 1334). SG acknowledges support from the F.R.S.-FNRS.
\end{acknowledgements}

\end{document}